\documentclass[preprint,aps,showpacs,preprintnumbers,amsmath,amssymb,superscriptaddress,nofootinbib,natbib,floatfix,11pt]{revtex4-1}
\usepackage{graphicx}
\usepackage{dcolumn}
\usepackage{placeins}
\usepackage[hidelinks]{hyperref}
\usepackage{xcolor}
\usepackage{booktabs}
\usepackage{multirow}
\usepackage{titlesec}
\usepackage{tikz}
\usepackage{tikz-3dplot}
\usetikzlibrary{arrows.meta, calc, positioning, fit}
\titleformat{\paragraph}
{\normalfont\normalsize\bfseries}{\theparagraph}{1em}{}
\titlespacing*{\paragraph}
{0pt}{3.25ex plus 1ex minus .2ex}{1.5ex plus .2ex}
\usepackage{xcolor}
\usepackage[title]{appendix}
\usepackage{soul}
\usepackage{siunitx}
\usepackage{enumitem}
\usepackage{physics}
\usepackage{booktabs}
\usepackage{subcaption}
\usepackage[footnote]{nicematrix}
\usepackage{pifont}
\usepackage[font=scriptsize]{caption}

\begin{document}
\newcommand{\AP}[1]{\textcolor{black}{#1}}
\newcommand{\APupdate}[1]{\textcolor{green}{#1}}
\newcommand{\SSupdate}[1]{\textcolor{blue}{#1}}

\title{
 A first-principles study of bcc chromium beyond the generalized gradient approximation (GGA)}
\author{Alma Partos}
\email{partos@doktorant.umk.pl}
\affiliation{Institute of Physics, Nicolaus Copernicus University, 87-100 Toru\'n, Poland}
\author{Igor Di Marco}
\affiliation{Institute of Physics, Nicolaus Copernicus University, 87-100 Toru\'n, Poland}\affiliation{Department of Physics and Astronomy, Uppsala University, Uppsala 751 20, Sweden}
\author{Shivalika Sharma}
\email{shivalikasharma26@gmail.com, shivalika@kentech.ac.kr}
\affiliation{Institute of Physics, Nicolaus Copernicus University, 87-100 Toru\'n, Poland}
\affiliation{Center for Theoretical Physics of Complex Systems, Institute for Basic Science (IBS), Daejeon 34126, South Korea}
\date{December 2025}
\renewcommand{\abstractname}{Abstract}
\begin{abstract}
\noindent The study of magnetism in transition metals is a cornerstone in understanding complex electronic and magnetic interactions in condensed matter systems. Among transition metal elements, body-centered cubic (bcc) chromium stands out because of its spin-density wave (SDW) ground state, posing a long-standing challenge for density functional theory (DFT). Conventional functionals, such as the generalized-gradient approximation (GGA) and the local-density approximation (LDA), fail to predict this experimentally observed incommensurate SDW as the ground state. In this study, we present a comprehensive DFT analysis of bcc Cr employing GGA and a variety of meta-GGA functionals. We evaluated total energies, structural parameters, and magnetic properties across a wide range of SDW wave vectors. Our results show that all meta-GGA functionals overestimate the local magnetic moments and enhance the nodal magnetic frustration, destabilizing the SDW state relative to the commensurate antiferromagnetic (AF) configuration. Tao--Perdew--Staroverov--Scuseria (TPSS) yields results closest to those of the GGA, thus providing the most adequate description of bcc Cr among the meta-GGA functionals. These results emphasize the need for the further development of non-local or hybrid functionals tailored for complex magnetic systems.
\end{abstract}
\maketitle
\section{Introduction}
\noindent Understanding the magnetic and electronic properties of materials is a central problem in modern science. Density functional theory (DFT) has become one of the most important tools in condensed matter physics, materials science, and quantum chemistry to predict the electronic structure and properties of materials from  first principles, using the electron density as the fundamental variable \cite{moruzzi1986,kohn1965,hohenberg1964}. Over the years, numerous exchange--correlation (XC) functionals have been developed within DFT to improve on the description of exchange and correlation for systems with varying degrees of complexity.
These functionals are commonly classified through a Jacob’s ladder, starting with the local-density approximation (LDA), followed by the generalized-gradient approximation (GGA), up to the more advanced meta-generalized gradient approximation (meta-GGA) and hybrid functionals \cite{perdew2001}. Despite these advances in XC functionals, DFT still struggles with materials exhibiting complex magnetic behavior, strongly correlated systems, Mott insulators, \emph{etc.}~\cite{cohen2012,jones2015,RevModPhys.78.865_kotliar}. One such case is bcc Cr, where experimental studies using neutron diffraction have revealed an incommensurate SDW as the ground state \cite{shull1953,fawcett}. A SDW is a spatially periodic modulation of the magnetization density, which in bulk Cr manifests as alternately-oriented magnetic moments on the corner and center atoms of the bcc structure. At low temperatures, the SDW is longitudinal, where the magnetic moments are aligned parallel to the wave vector, $\vec{q}$ [Fig.~1(c)]. At a temperature of $T_{\text{SF}} \simeq 123$ K \cite{fawcett}, it undergoes a spin-flip transition to a transverse-wave state, where the magnetic moments rotate perpendicularly to the direction of $\vec{q}$ [Fig.~1(b)]. Above $T_{\text{N}} \simeq 311$ K, the system becomes paramagnetic \cite{fawcett,overhauser1962}. The SDW period is determined by the magnitude of the wave vector $\vec{q} = \frac{2\pi}{a}(1 - \delta)\hat{n}$ \cite{fawcett}, where $\delta$ is the incommensurability parameter, $a$ is the lattice constant, and $\hat{n}$ is a unit vector along the propagation direction of the wave. $\delta$ measures the deviation from the simple commensurate AF configuration [Fig.~1(a)] in which $\delta = 0$. Experimental studies have reported $\delta \simeq 0.048$, which corresponds to a period of roughly $21$ conventional unit cells \cite{shull1953,fawcett}. The stability of this incommensurate SDW was first explained by Overhauser \cite{overhauser1962}, who proposed that it arises due to Fermi surface nesting. Various theoretical studies based on DFT, employing LDA and GGA functionals, along with different electronic-structure methods such as projected augmented wave (PAW), full-potential linear augmented wave (FLAPW), and linear muffin-tin orbital (LMTO), have predicted that the commensurate AF state is more stable than the SDW state \cite{hafner,Cottenier,marcus1998b,chen1988,singh1992,bihlmayer2001,barreteau}. Beyond standard DFT, studies using the $\text{DFT} + U$ approach have also been used to investigate the magnetic ground state of bcc Cr \cite{vanhoof_cottenier_nodonModel,bisti_evidence_2025}. However, this only improved the agreement between the theoretical and experimental magnetic moments, yet still predicted the AF order as the stable state over the SDW. To account for this discord with the experiment, the authors of Ref.~\cite{uzdin2006}, and subsequently those of Ref.~\cite{vanhoof_cottenier_nodonModel}, put forward a different perspective: that the SDW is not the true ground state. Instead, they suggested that the SDW arises from small thermal excitations called ``nodons", which act on a state of commensurate AF order \cite{vanhoof_cottenier_nodonModel,uzdin2006}. Through this model, they emphasized that the AF configuration is the actual ground state, while the observed SDW emerges as a low-energy excitation. However, this physical picture has clashed with recent experimental evidence reporting various signatures of the SDW state in the angle-resolved photoemission spectrum (ARPES) \cite{bisti_evidence_2025}. Despite extensive prior research, none of the DFT-based approaches to date have succeeded in predicting the SDW as the true ground state of bcc chromium.
 Possible reasons for this failure include: (1) the semilocal nature of GGA and LDA functionals, which are unable to capture the non-local exchange--correlation effect of complex magnetic structures like SDWs, an issue also addressed by Capelle and Oliveira, who pointed out that certain non-collinear or incommensurate magnetic states may only be stabilized by a non-local exchange--correlation potential \cite{Capelle}; and (2) the presence of strong correlation effects, which have been shown to play a key role for magnetic $3d$ elements \cite{PhysRevLett.87.067205_lichtenstein_katsnelson_2001,hausoel_local_2017,PhysRevB.79.115111_DiMarco_2009,PhysRevB.85.205109_SanchezBarriga,di_marco_-mn_2009,PhysRevB.87.115130_Pourovskii,pourovskii_electronic_2020}.\\
 \begin{figure}
    \centering
\includegraphics[width=\linewidth]{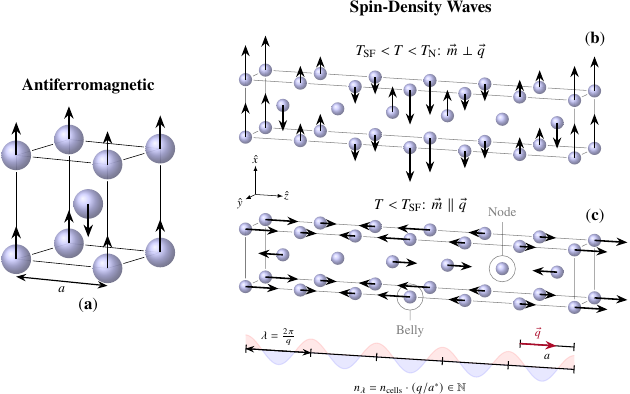}
    \caption{(a) c-AF configuration of the bulk-bcc Cr. (b) The transverse SDW state, with magnetic moments (here of an arbitrary maximal magnitude) perpendicular to the wave vector ($\vec{q}$), occurring between $T_{\text{SF}}< T< T_{\text{N}}$. (c) The longitudinal SDW state ($T < T_{\text{SF}}$), which is our main subject of study. Encircled are a ``node'' atom, characterized by a zero magnetic moment, and a ``belly'' atom, in turn characterized by the maximal magnetic moment. Below is an illustration of the incommensurate modulation wave with wave length $\lambda = 2\pi/q = a/(1 - \delta)$, where $q = \abs{\vec{q}}$ and $\vec{q} = \frac{2\pi}{a}(1 - \delta) \, \hat{z}$. Here, $n_{\lambda}$ is the number of modulation wave lengths and $n_{\text{cells}}$ the number of bcc-Cr cells such that the SDW period fits exactly in the supercell: $n_{\text{cells}}a = n_{\lambda}\lambda$.}
    \label{fig:spin configs}
\end{figure}
\noindent To overcome the limitations of conventional functionals, a class of more advanced meta-GGA functionals has been developed. These incorporate an additional semilocal ingredient, the kinetic energy density, to better account for inhomogeneous electronic environments and to partially reduce self-interaction errors \cite{metagga_scan_prediction,meta_gga,meta-gga2}. TPSS (Tao--Perdew--Staroverov--Scuseria) is a widely used meta-GGA functional that satisfies several exact constraints and has proven successful in improving atomization energies, molecular geometries, and weak interactions \cite{TPSS,assessingtheperformance_tpss}. The SCAN (strongly constrained and appropriately normed) functional is another non-empirical meta-GGA that, compared to TPSS, satisfies a larger set of exact constraints \cite{PhysRevB111}, and captures intermediate-range van der Waals interactions. It has been successfully applied to various systems including semiconductors, strongly correlated oxides, layered materials, and van der Waals solids \cite{Fu_2019}.
A reformulated version of SCAN is SCAN-L \cite{deorbitalization_strategies,Mejia_Rodriguez_2018}, a deorbitalized variant that improves numerical stability and reduces the overestimation of magnetic moments in itinerant magnets \cite{shortcomings_of_metaGGAs_magnetism,analysis_of_overmagn,Fu_2019}. In parallel, the semi-empirical M06-L (Minnesota 2006, local) functional \cite{zhao_new_2006} includes parameters fitted to thermochemical and kinetic data, and shows good performance for molecular systems, main-group thermochemistry, and non-covalent interactions \cite{peverati2012ML06}. These meta-GGA functionals have shown notable success for a range of materials classes, including semiconductors, magnetic systems, and correlated oxides \cite{TPSS,MO201738,true_ground_state_of_intermetallic,assessingtheperformance_tpss}.
To date, no comprehensive study has systematically examined the performance of meta-GGA functionals for the magnetic phases in bcc Cr. This is surprising, since Cr stands out from the other transition metals due to its critical magnetic behavior that is not well described by LDA or GGA. For example, experiments reveal a significant disparity between the magnetic moment at the surface ($\sim 2.96$ $\mu_{\text{B}}$) and in the bulk ($\sim 0.6$ $\mu_{\text{B}}$) \cite{Zabel_1999}. More unusual behavior at the Cr (001) surface was reported in scanning tunneling microscopy (STM) experiments, which was interpreted in terms of orbital Kondo resonance or surface states \cite{kolesnychenko_real-space_2002,PhysRevB.72.085453_Hanke,PhysRevB.77.233409_budke,PhysRevB.96.245137_peters}. This sensitivity to dimensionality highlights the importance of selecting an exchange--correlation functional with the best possible accuracy. Therefore, a detailed analysis of the performance of meta-GGA functionals in predicting structural and magnetic properties across different magnetic phases of Cr is needed in order to understand the complex magnetic behavior of the system.\\
\noindent In this work, we aim to investigate whether any meta-GGA functionals, in comparison to conventional GGA, can predict the SDW as the magnetic ground state over the AF phase.
We present a comparative theoretical study using a set of meta-GGA functionals, including SCAN, SCAN-L, TPSS and M06-L, along with the PBE-GGA functional, and evaluate the total energies of both the AF and SDW magnetic phases to determine the correct ground state. 
We also analyze the structural, magnetic, and electronic ground-state properties predicted by these functionals to describe the magnetic behavior of bcc Cr. The paper is organized as follows: Section II describes the computational Methodology; Section III presents the Results and Discussion, including subsections on structural optimization and the assessment of the meta-GGA functionals for bulk Cr with commensurate AF and incommensurate SDW phases, followed by electronic structure analysis of the latter; Section IV covers the Conclusion.
For convenience, hereafter we use the acronyms ``c-AF'' and ``SDW'' to denote the commensurate antiferromagnetic and incommensurate SDW phases, respectively.
\section{Methodology}
\noindent Structural optimization of bulk bcc Cr in the c-AF and SDW phases was performed using the PAW \cite{blochl_PhysRevB.50.17953_1994,kresse_PhysRevB.59.1758_1999} method, as implemented in the \textsc{Vienna Ab-Initio Simulation Package} (VASP) \cite{kresse_PhysRevB.47.558_1993,kresse_PhysRevB.49.14251_1994,kresse_PhysRevB.54.11169_1996, kresse_cms.6.15_1996}.
  The Perdew--Burke--Ernzerhof (PBE) functional was used, in its standard parameterization, for the GGA calculations, while TPSS, M06-L, SCAN, and SCAN-L were employed for meta-GGA calculations \cite{perdew_PhysRevLett.77.3865_1996,meta_gga, meta-gga2,TPSS,PhysRevLett-metagga}. 
The $3d$ and $4s$ electrons were treated as valence states. The ionic relaxation was performed with fixed cell volume and shape, allowing only the atomic positions to relax, using a force convergence criterion of $10^{-3}$ eV/\AA. The total energy convergence criterion for the self-consistent cycle was set as $10^{-7}$ eV. Based on our convergence tests, the plane-wave cutoffs were set to $350$ eV for the GGA and $550$ eV for all meta-GGA calculations excepting ones involving the SCAN-L functional. \AP{For SCAN-L, higher plane-wave cutoffs and dense FFT grids were required for numerical stability, a fact which may be related to known sensitivities of both the base SCAN functional and deorbitalized Laplacian-level meta-GGAs}  \cite{PhysRevB.102.121109_metagga_performance_in_solids,assessing_the_performance_PhysRevB.108.024306,accurate_amd_numerically_efficient_r2scan_doi:10.1021/acs.jpclett.0c02405, VASP_METAGGA_Wiki}. Based on \AP{our} convergence tests and established guidelines \cite{VASP_METAGGA_Wiki}, we adopted a cutoff of 700 eV, for which all relevant energy differences are well converged. For all functionals, the Brillouin zone was sampled using a conventional Monkhorst-Pack grid of $12 \times 12 \times 12$ and $12 \times 12 \times 1$ \textbf{k} points for the c-AF and SDW phases, respectively.
We have used a Gaussian smearing, with a smearing parameter of $\sigma = 0.01$ eV. For the meta-GGAs, we additionally took into account the aspherical corrections from the gradient corrections inside the PAW sphere \cite{VASP_METAGGA_Wiki}, which need not be accommodated in the GGA formulation.\\
The subsequent analysis of the electronic structure was limited to PBE, TPSS and SCAN-L. This set of meta-GGA functionals was selected for \AP{detailed} self-consistent calculations (SCF) based on the analysis of magneto-structural properties, as discussed in Section III. The basis set and Brillouin zone sampling were kept consistent with those used during structural optimization both for the bcc Cr c-AF and for SDW supercells. In a few cases, final self-consistent calculations exhibited residual forces up to $\sim 3 \times 10^{-3}$ eV/Å, which do not affect the reported energy differences. For density of states (DOS) calculations, the sampling \textbf{k}-mesh was refined to $24 \times 24 \times 24$ for the bulk c-AF and $24 \times 24 \times 1$ for the SDW supercells. Gaussian smearing was used, with a width of $\sigma = 0.01$ eV for the SCF and $\sigma = 0.1$ eV for the DOS calculations. Spin-orbit coupling was not taken into account, as its impact on the electronic and magnetic properties of the SDW \AP{was} proved \AP{to be} negligible in previous DFT studies \cite{hafner, vanhoof_cottenier_nodonModel}.

\section{Results and Discussion}
\subsection{Structural optimization and assessment of meta-GGA functionals}
\subsubsection{Commensurate AF phase}
\noindent We first optimized the lattice parameter of bulk bcc Cr by performing calculations for the non-magnetic (NM) and c-AF configurations [Fig.~1], using both GGA and meta-GGA functionals. Since bcc Cr does not possess internal structural degrees of freedom, the ionic positions remained fixed, and no atomic relaxation was needed. The equilibrium lattice constant was determined by evaluating the total energies across a range of lattice constants, also including the experimental value $a_{\text{exp.}} = 2.884$~\AA~\cite{hafner}. Note that this experimental lattice constant corresponds to an SDW ground state, rather than an AF or NM configuration; however, previous studies have pointed out that the AF and SDW states have similar equilibrium lattice constants \cite{hafner,Cottenier}. Initially, we adopted one recommended approach for converging the meta-GGA calculations \cite{VASP_METAGGA_Wiki}, which consists in the usage of pre-converged GGA wavefunctions as the starting point for the SCF cycle \cite{VASP_METAGGA_Wiki,kingsbury_PhysRevMaterials.6.013801,PhysRevB111,assessingtheperformance_tpss}. This was done with the \AP{purpose} of facilitating the forthcoming calculations of larger supercells. This strategy did consistently produce converged energies, but struggled for small lattice constants ($a\leq2.79$~\AA), for which the system relaxed to the NM state for all meta-GGA functionals. This tendency is due to the fact that the pre-converged GGA wavefunctions correspond to the NM state in that regime, which produces discontinuities and unphysical trends in the energy and magnetic moment curves for the meta-GGAs. This is illustrated in Figures S1 and S2 in the Supplemental Material (SM). To address this issue, we repeated all the meta-GGA SCF calculations without using pre-converged GGA wavefunctions, starting from
scratch. The system then consistently converged to the AF state, yielding smooth energy vs.~lattice constant curves and more physically reasonable magnetic moments. Magnetic moment and energy curves
obtained with these two strategies are perfectly consistent for lattice constants above $a \simeq 2.79$~\AA, which confirms the stability of our calculations. Overall, this behavior is in line with the general sensitivity of meta‑GGA functionals to initial conditions and the numerical convergence challenges encountered in marginally magnetic systems \cite{PhysRevB111}. The resulting energy vs.~lattice constant curves are presented in Figure 2. For each functional, we observe that the energy curves of the NM and AF configurations overlap up to a certain value of $a$, which marks the onset of the formation of a finite local magnetic moment. As soon as a local magnetic moment forms, the AF state becomes energetically more favorable. Compared with GGA, the meta-GGA functionals show the NM and AF energy curves splitting at significantly smaller values of the lattice constant. Among them, M06-L yielded the earliest split at $a=2.58$~\AA. This shows the more excessive tendency of the meta-GGA exchange--correlation functionals to stabilize magnetic ordering and is in line with previous studies on Fe, Co and Ni reporting overestimated magnetic moments \cite{metagga_scan_prediction,shortcomings_of_metaGGAs_magnetism}. The tendency to overestimate magnetism is also evident in the magnetic moment curves as a function of lattice constant, shown in Figure 3. SCAN and M06-L exhibit unrealistically large magnetic moments compared to SCAN-L, TPSS, GGA, and, most importantly, experimental values.
\begin{figure}
    \centering
    \includegraphics[width=\linewidth]{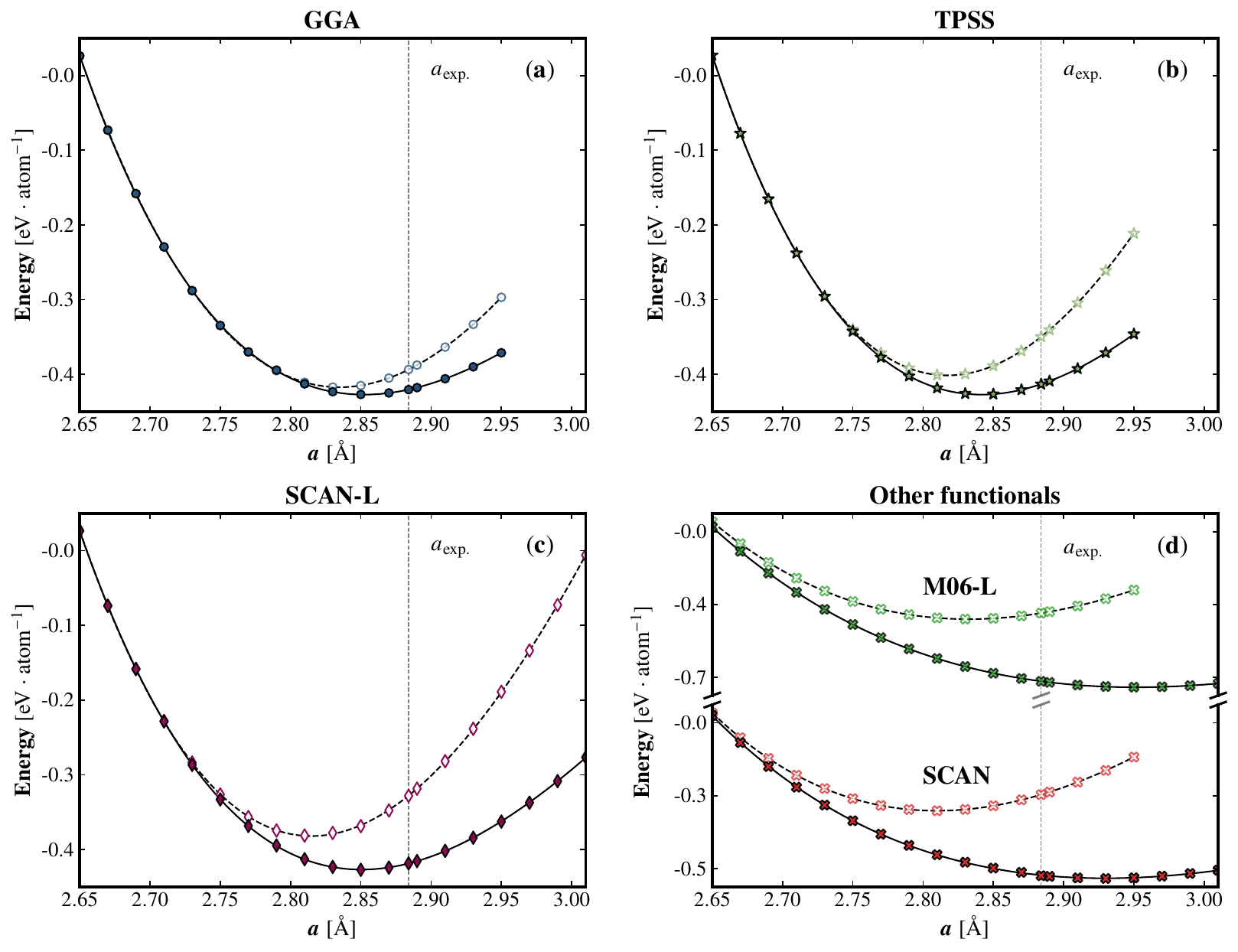}
\footnotesize\caption{Total energy as a function of the lattice constant for NM (open symbols) and c-AF (filled symbols) bulk-bcc Cr calculated using (a) GGA, and the meta-GGAs (b) TPSS, (c) SCAN-L, (d) M06-L and SCAN, as implemented in VASP. Vertical dashed lines mark the experimental lattice constant ($a_{\text{exp.}}$). All curves are plotted using a constant offset determined by the NM energy at the starting point of the range, i.e.~$a=2.65$ \AA.}
    \label{fig:E_vs_latt}
\end{figure}
\begin{figure}
    \centering
    \includegraphics[width=0.5\linewidth]{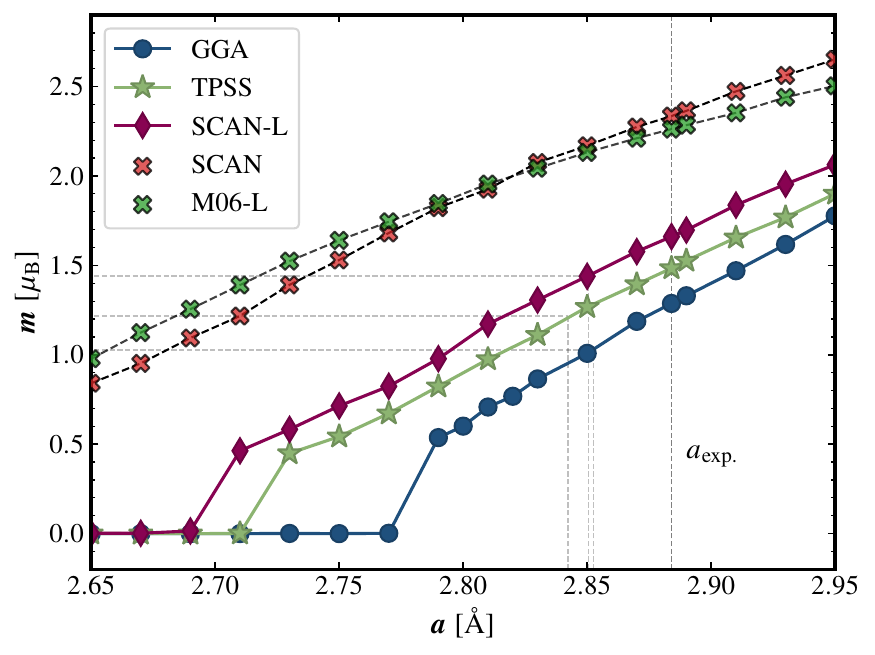}
    \caption{Local magnetic moment as a function of the lattice constant ($a$) for bulk-bcc Cr in the c-AF phase, obtained using the GGA, TPSS, SCAN-L, M06-L and SCAN functionals, as implemented in VASP. The dashed lines indicate the magnetic moments calculated at the equilibrium lattice constants.}
    \label{fig:magmom_vs_latt}
\end{figure}
\begin{table}[b]
    \centering
    \caption{Difference, $\Delta E_{\text{ref.}}$, between the minimum total energy per atom in the reference lower-energy state, which is c-AF throughout, and the minimum total energy per atom in the considered system; we also show the equilibrium volume ($V_0$), bulk modulus ($B$), and equilibrium lattice constant ($a_{\text{theo.}}$). All quantities are determined from the Birch-Murnaghan equation of state for the bulk-bcc Cr in both NM and AF configurations. Calculations were performed using the PAW method as implemented in VASP, with GGA and meta-GGA (TPSS, SCAN, SCAN-L, and M06-L) XC functionals. The magnetic moments provided for the AF configuration are calculated at both the experimental and theoretical lattice constants. The experimental values are included for comparison.}
    \setlength{\tabcolsep}{6pt}
    \begin{NiceTabular}{c cS[round-mode=places, round-precision=2]S[round-mode=places, round-precision=2]S[round-mode=places, round-precision=0, table-format=4.0]S[round-mode=places, round-precision=2]S[round-mode=places, round-precision=2]S[round-mode=places, round-precision=2]}
   \CodeBefore
    \rowlistcolors{2}{white, gray!10}[restart,cols={2-8}]
    \Body
    \toprule[1pt]
    \Block[draw=black, line-width=1pt]{13-8}{}
       &  & & & & & \multicolumn{2}{c}{$m$ [$\mu_{\text{B}}$]} \\
      \cmidrule[1pt](lr){7-8}
       \textbf{System} &  $\boldsymbol{E_{\text{xc}}}$& {$\Delta E_{\text{ref.}}$ [$\text{eV} \cdot \text{atom}^{-1}$]} & {$V_0$ [$\text{Å}^3$]} & {$B$ [GPa]} & {$a_0$ [Å]} & {$m(a_{\text{exp.}}) $\tabularnote{The magnetic moment calculated at the experimental lattice constant, $a_{\text{exp.}} = 2.884$ Å \cite{hafner}.} }& {$m(a_{\text{theo.}})$} \\
    \midrule[1pt]
    \Block[draw=black, line-width=0.6pt]{11-1}{}
    \Block[draw=black, line-width=0.6pt]{11-2}{}

        \multirow{5}{*}{\textbf{NM}} & GGA & 0.009671 & 22.783655 &  261.281782 & 2.835 &       & \\
         & TPSS & 0.022755 & 22.361838 & 280.198204 & 2.817 &       & \\
        & SCAN-L & 0.040372 & 22.304597 & 280.229776 & 2.8149117979954585 &     &     \\
        & SCAN & 0.238569 & 22.182260 & 284.791977 & 2.80975592 &     &     \\
        & M06-L & 0.30842 & 22.723179 & 271.156659 & 2.8324116 &     &     \\
        \cmidrule[0.6pt](lr){1-2}
        \multirow{5}{*}{\textbf{AF}}  & GGA &  {0} & 23.209329 & 179.624773 & 2.852 & 1.287 & 1.026 \\
        & TPSS & {0} & 22.943997 & 187.564542 & 2.841556925323064 & 1.484 & 1.218   \\
        & SCAN-L & {0} & 23.159807 & 174.640518 & 2.850438313996444 & 1.661 & 1.440   \\
        & SCAN & {0} & 25.148545 & 118.062525 & 2.929797599652859 & 2.340 & 2.653   \\
        & M06-L & {0} & 25.719798 & 131.436152 & 2.9518153170643853 & 2.262 & 2.504   \\
        \midrule[0.3pt]
        \midrule[1pt]
        \textbf{SDW} & \textbf{Expt}.\tabularnote{The experimental values are taken from Refs.~\cite{hafner,marcus1998b, Cottenier}.}& &   & 191 &  2.884 & \multicolumn{2}{c}{0.62}\\
        \bottomrule[1pt]
    \end{NiceTabular}
    \label{tab:properties_2}
\end{table}
\noindent Furthermore, using the Birch-Murnaghan equation of state \cite{ase-paper}, theoretical equilibrium lattice constants, volumes, and bulk moduli were extracted. A summary of the predicted physical properties is shown in Table I. The GGA shows good consistency with previously reported theoretical data for lattice constants, bulk moduli, and magnetic moments \cite{hafner, barreteau}. Focusing on the NM state, all meta-GGA functionals provide results similar to those of the GGA. When spin-polarization is allowed, however, only TPSS and SCAN-L yield values that closely agree with those obtained using GGA.
In contrast, SCAN and M06-L exhibit marked differences in equilibrium lattice constants, bulk moduli and magnetic moments. If we focus on the c-AF state, which should offer a good comparison with the experimental values, the data reported in Table I demonstrate that the meta-GGA functionals SCAN and M06-L both predict unphysical structural and magnetic properties for Cr, in agreement with previous studies on Fe, Co, and Ni, as well as other transition metal compounds exhibiting itinerant magnetism \cite{analysis_of_overmagn,shortcomings_of_metaGGAs_magnetism,true_ground_state_of_intermetallic}. This behavior stems from the tendency of SCAN to enhance exchange splittings relative to PBE-GGA, leading to the formation of exaggerated magnetic moments.
The marked improvement observed with SCAN-L further suggests that the overmagnetization can be remedied through the deorbitalization of SCAN \cite{deorbitalization_strategies} also for bcc Cr.
 Although GGA, TPSS, and SCAN-L also overestimate magnetic moments relative to experiment, they yield comparatively accurate lattice constants and bulk moduli. Therefore, these XC-functionals can be considered as more reliable for describing the electronic and magnetic properties of Cr. We proceed with these functionals for all subsequent electronic structure calculations of the SDW phase and the corresponding data analysis.

\subsubsection{Incommensurate SDW phase}
\noindent Following the optimization of the c-AF phase, we constructed a series of supercells to model the SDW states propagating along the
$z$ direction. For example, the SDW with a wave vector $\vec{q}/a^* = 11/12\,\hat{z}$, where $a^*=2\pi/a$, corresponds to a supercell of $1\times1\times12$, containing 24 Cr atoms [\emph{cf.}~Fig.~1(c)]. 
The modulation vector was expressed, \emph{e.g.}, as $\vec{q}/a^* = (0,0,11/12)=(0,0,0.916)$, such that $q = \abs{\vec{q}}$ refers to the scalar component of the wave vector along its $\hat{z}$ axis. We considered a set of wave vectors, $q/a^* \in \{11/12, 13/14, 17/18, 19/20, 21/22 \} $, with $q/a^* = 19/20 = 0.95$ being closest to the experimental value \cite{fawcett}.  
The ionic relaxation was performed for each supercell using the theoretical lattice constants obtained from the optimization functionals discussed above, which are listed in Table I. This choice is justified by the negligible difference in equilibrium lattice constants between c-AF and SDW states that has previously been reported in the literature \cite{Cottenier}.
Following our previous discussion on meta-GGA convergence strategies, no problems are to be expected for lattice constants larger than $a = 2.79$ Å, where the results are robust. Therefore, for the SDW calculations, where all theoretical lattice constants lie above this value, we employed pre-converged GGA wavefunctions as the initial starting point. 
This method led to reliable convergence at a reduced computational cost \cite{VASP_METAGGA_Wiki}.
The relaxation caused the interlayer spacing to contract around ``node" and expand near ``belly" atoms. Here, a node atom refers to a site where the magnetic moment is nearly zero, while a belly (anti-node) atom refers to a site hosting the overall highest (or almost-highest) magnetic moment, as illustrated in Figure 1(c).
Hafner \emph{et al.}~reported contractions of $-0.30$~\% around the nodes and expansions of 0.18~\% at the antinodes for the GGA functional \cite{hafner}. They concluded that these changes had little effect on the total energy and that the higher stability of the AF state with respect to the SDW state should not be affected by them. In contrast, our results show larger displacements, particularly for meta-GGA functionals (TPSS, SCAN-L) that exhibit contractions of $-0.61$ and $-0.84$~\% and expansions of $0.36$ and $0.31$~\% for the smallest supercell with $q/a^*=11/12$, as summarized in Table SI of the SM. Nevertheless, in agreement with Hafner \emph{et al.}'s conclusion, these larger displacements do not lead to a change of ground state, as illustrated in the next section.
\begin{figure}[t]
    \centering
    \includegraphics[width=\linewidth]{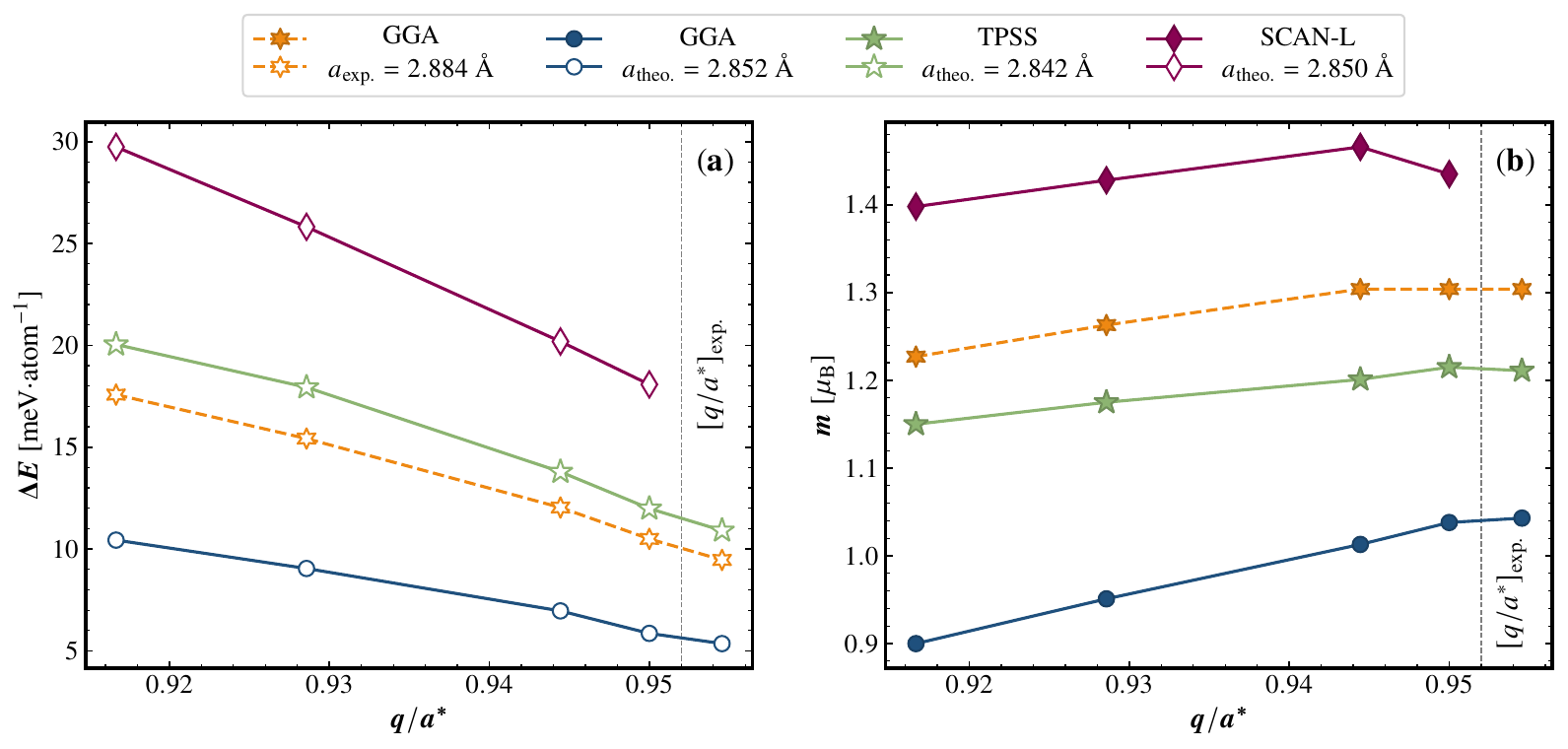}
    \caption{(a) Energy difference $\Delta E= E_{\text{SDW}}-E_{\text{AF}}$ between the SDW, for all tested wave vectors, and the c-AF state. (b) Magnetic moment of the belly atom (SDW amplitude) as a function of the modulus of the SDW ordering vector.}
    \label{fig:deltaE}
\end{figure}
\subsection{Magnetic ground state and electronic structure analysis of the SDW phase}
\noindent We performed the electronic structure calculations for the optimized SDWs using different functionals (GGA, TPSS and SCAN-L) along the same range of wave vectors. For each functional, the calculations were carried out at the corresponding theoretical lattice constant, as listed in Table I. In the case of the GGA, we also considered the experimental lattice constant to allow for comparison with previous literature \cite{hafner,Cottenier}. Figure 4(a) shows the energy difference between the SDW and AF configurations as a function of $q/a^*$. For the SCAN-L functional, calculations at $q/a^* = 21/22$ and above were not performed due to the comparatively high computational requirements.\\
\noindent For all functionals, at any given $q/a^*$, we find that the energy difference ($\Delta E = E_{\text{SDW}} - E_{\text{AF}}$) is always positive, which indicates that the AF state is energetically favorable over the SDW. In all cases, the curve $\Delta E$ shows an almost linear decrease with increasing $q$. This is expected since the AF state corresponds to the SDW state with $q/a^*=1$. For the GGA, at $a_{\text{theo.}}$ as well as $a_{\text{exp.}}$, these results are in qualitative and quantitative agreement with previous studies \cite{hafner}. For example, the GGA($a_{\text{exp.}}$) calculation for $q/a^*=17/18$ gives $\Delta E \sim 12$ meV/atom, which closely matches the value reported by Hafner \emph{et al.}~under the same conditions \cite{hafner}. Overall, the smallest obtained energy difference occurs for GGA($a_{\text{theo.}}$) at $q/a^* = 21/22$ and amounts to circa $5$ meV. In contrast, the meta-GGA functionals SCAN-L and TPSS yield significantly larger $\Delta E$ values [Fig.~4(a)].
 \begin{figure}
    \centering
    \includegraphics[width=1\textwidth]{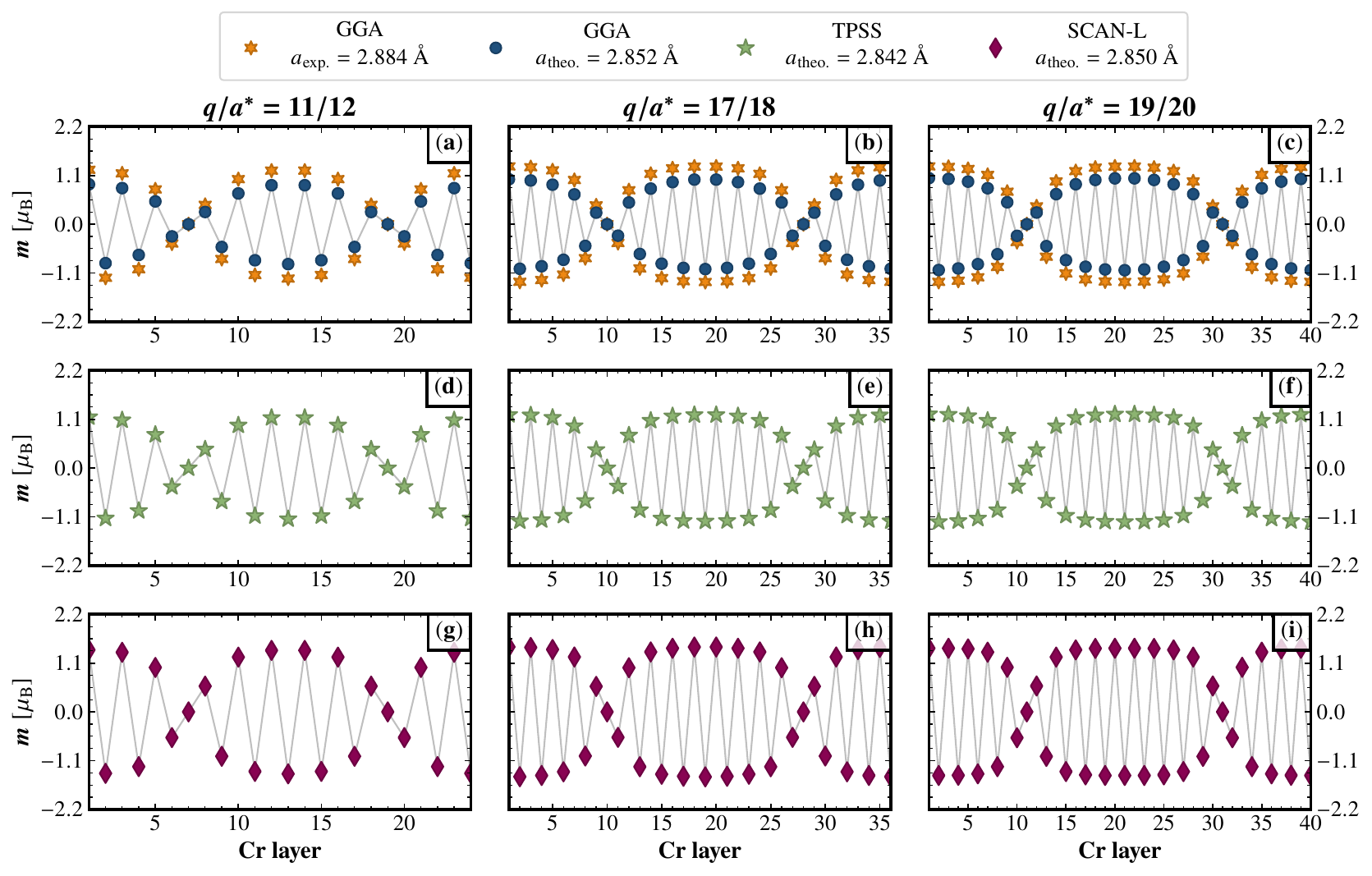}
    \caption{Local magnetic moments for SDWs with $q/a^{*} \in \{11/12, 17/18, 19/20 \}$ as calculated using the theoretical lattice constants predicted by each functional. For the GGA, values corresponding to the experimental lattice constant are included for comparison. The atoms are labeled so that consecutive indices refer to nearest-neighbor corner- and center atoms, equivalent to following the $(111)$ direction in the bcc lattice.}
    \label{fig:magmom_sdw}
\end{figure}
 \noindent 
 This may arise as a result of the tendency of these functionals to favor stronger local magnetic moments. What is clear from Figure 4(a) is that $\text{GGA}(a_{\text{theo.}})$ provides the lowest values of $\Delta E$ and thus offers the highest likelihood for the SDW to be stabilized by alternative mechanisms, such as those discussed in the Introduction section. The main result emerging from our data is that currently available meta-GGA functionals do not capture the microscopic mechanisms stabilizing the SDW state.\\
  \noindent To explore whether magnetic properties help explain this trend, we analyzed the local magnetic moment of the belly atom (the Cr atoms at the centers of the SDW peaks) as a function of $q/a^{*}$ [Fig.~4(b)], as well as the magnetic moment profiles [Fig.~5], for three selected wave vectors across all functionals. Figure 4(b) shows that all functionals overestimate the magnetic moment compared to the experimentally reported largest moment in the SDW ($\sim0.62$ $\mu_{\text{B}}$) \cite{fawcett}. For any given functional, the belly-atom moment increases with $q/a^{*}$, accompanied by an evolution from a triangular (nearly sinusoidal) to a more rectangular profile [Fig.~5]. At large $q/a^*$ values, the SDW belly atom approaches bulk AF order, with nearly uniform magnetic moments across most atoms and only localized reductions near the nodes. Such rectangular profiles minimize the energetic penalty associated with frustrated moments at the nodes, leading to a reduction of $\Delta E$. This behavior can be explained in terms of local exchange fields, and is consistent with the interpretation provided by Hafner \emph{et al.}~\cite{hafner}.
Interestingly, for meta-GGA functionals such as TPSS and SCAN-L, the transition to a rectangular-like magnetic profile occurs at smaller wave vectors ($q/a^*=17/18$) than in GGA, with much larger magnetic moments [Fig.~5].  This supports the idea that the overestimation of the exchange coupling and local magnetic moments by meta-GGA functionals amplifies the energy cost associated with quenching the magnetic moments at the nodes. Fixed-spin-moment calculations in the c-AF configuration, as shown in Fig.~S3 of the SM, appear to further substantiate this interpretation. This aspect is connected to the fact that meta-GGA functionals tend to predict a magnetic ground state in many non-magnetic solids such as Sc, V, and Pd \cite{shortcomings_of_metaGGAs_magnetism}. \AP{Overall}, a stronger exchange interaction makes it more energetically costly to deviate from the AF configuration, thereby destabilizing the SDW state relative to the AF state.
 Further insights are provided by the total density of states (TDOS), shown in Figure 6. For each functional, we compare the NM, AF and SDW at $q/a^* = 19/20$. For all functionals, the AF state exhibits a deep pseudogap at the Fermi energy ($E_{\text{F}}$). At $E_{\text{F}}$, the SDW state exhibits a spectral weight lying between those of the AF and NM states [see the insets of Fig.~6]. The larger TDOS of the SDW state with respect to the AF state reflects the nodal frustration discussed above and is consistent with previous DFT studies \cite{barreteau, hafner}. Among the functionals, the residual TDOS of the SDW increases from GGA to TPSS to SCAN-L, correlating with an increased $\Delta E$, and suppressing the stability of the SDW by over-penalizing the nodes \cite{PhysRevB.108.115156-pesudo}. Figure 7 shows the $3d$-projected DOS (PDOS) for the SDW configuration at $q/a^* = 19/20$. At the node sites, where the magnetic moment vanishes, the PDOS for all functionals exhibits NM-like behavior. A significant residual PDOS at the Fermi level is observed, which is expected to penalize the stability of this state. Morever, the residual PDOS at $E_{\text{F}}$ for the node atoms is larger for SCAN-L and TPSS than for PBE-GGA. At the belly sites, all functionals exhibit a clear exchange splitting, whose magnitude follows the relation $\text{GGA}(a_{\text{theo.}}) < \text{GGA}(a_{\text{exp.}}) < \text{TPSS}(a_{\text{theo.}}) < \text{SCAN-L}(a_{\text{theo.}})$. Thus, SCAN-L shows the largest exchange splitting compared to TPSS and GGA [Fig.~7]. We also note that the larger the exchange splitting, the smaller the PDOS around $E_{\text{F}}$ for the belly atoms. This demonstrates that the stronger exchange splitting affects the energy stability mainly through varying the degree of frustration at the node atoms. Overall, GGA with $a_{\text{theo.}}$ offers the best agreement with the available experimental evidence regarding energy trends and magnetic behavior \cite{hafner, Cottenier}. Nonetheless, all tested functionals fail to predict SDW as the true ground state, highlighting the need for further development in XC functionals to better capture complex spin modulations in itinerant magnets like bcc chromium.

\begin{figure}
    \centering
    \includegraphics[width=\linewidth]{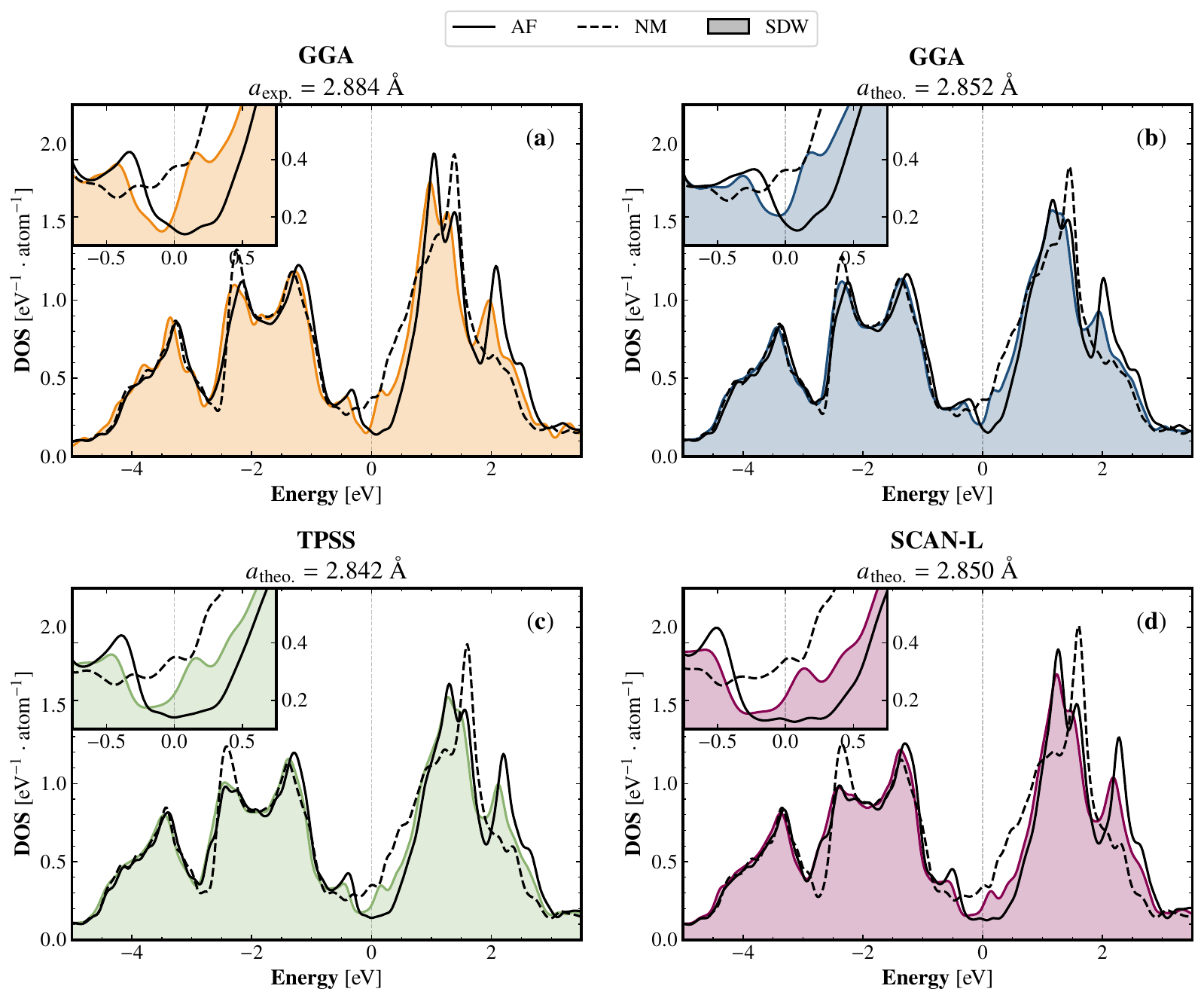}
    \caption{TDOS for the AF, NM, and SDW (at $q/a^* = 19/20$) configurations using GGA, TPSS, and SCAN-L at the theoretical, and for the GGA also experimental, lattice constants. The insets show a magnified region around the Fermi energy. The data have been normalized \AP{to one formula unit} for comparison.}
    \label{fig:tdos}
\end{figure}

\begin{figure}
   \centering
    \includegraphics[width=\linewidth]{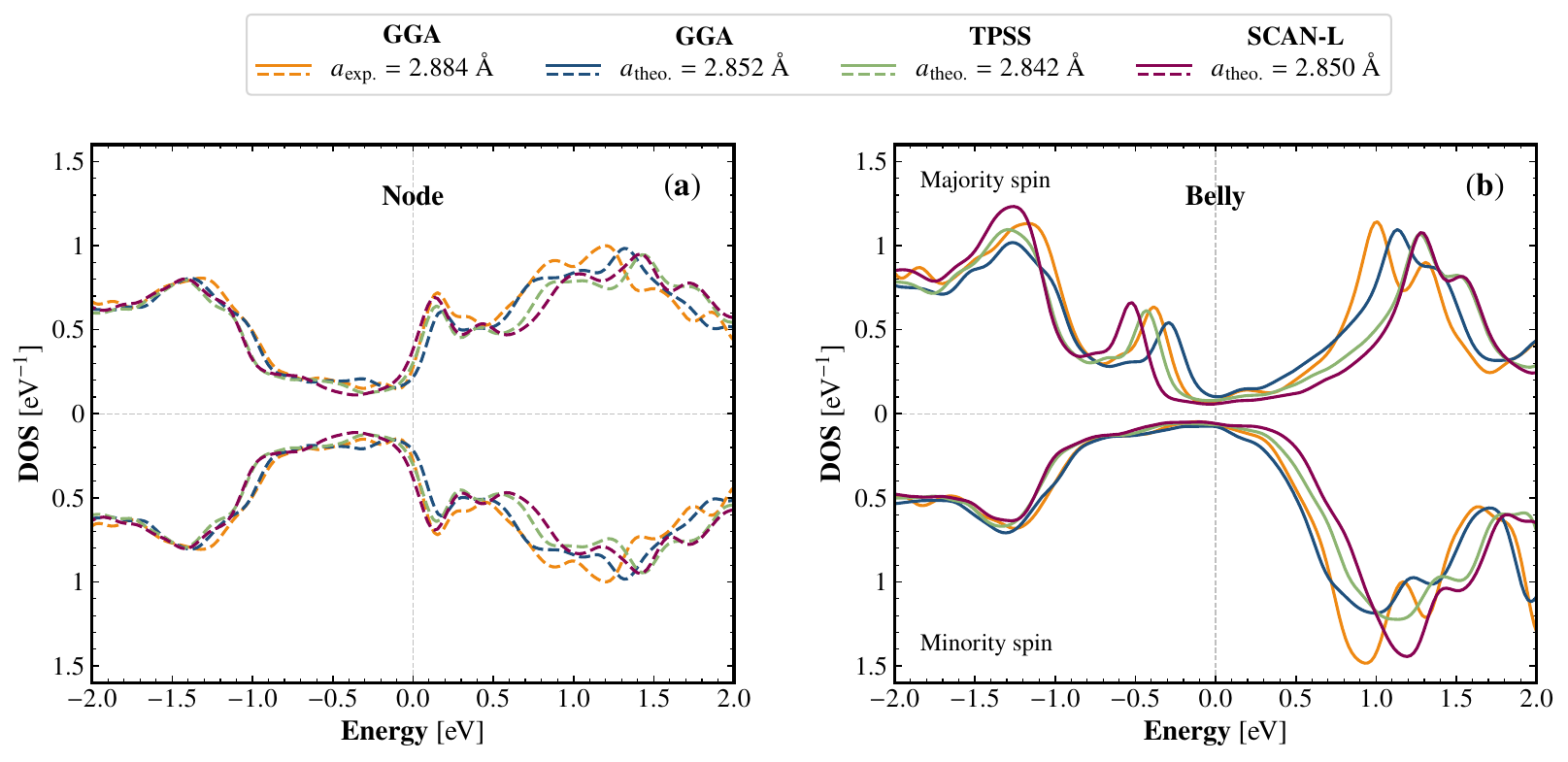}
    \caption{Spin-polarized PDOS for the Cr-$3d$ states at the node (dashed lines) and belly (solid lines) atom, respectively, for the SDW with $q/a^* = 19/20$.}
    \label{fig:pdos_v2}
\end{figure}

\section{Conclusion}
\noindent In this study, we performed a comprehensive theoretical investigation to determine whether meta-GGA functionals are capable of predicting the SDW as the magnetic ground state of bcc Cr. Our results show that none of the higher-order functionals tested provides a satisfactory description of the SDW phase, whereas PBE-GGA, despite being the simplest approximation, offers the most balanced performance in terms of magnetic moments, energy trends, and computational cost. These observations are consistent with earlier reports that demonstrated an improved accuracy of meta-GGAs for AF insulators, \emph{e.g.}~transition-metal oxides, but a poor description of \AP{magnetic} metals \AP{in general} \cite{PhysRevB.100.035119_varignon,PhysRevB.105.165111_varignon_bandGaps,shortcomings_of_metaGGAs_magnetism,PhysRevB111}. In a very recent article, Desmarais \emph{et al.}~\cite{PhysRevLett.134.106402_desmarais_mSCAN} connected this problem to the lack of form invariance under $\rm U(1) \times \, SU(2)$ gauge transformations, and suggested a new functional, mSCAN, that may potentially outperform other meta-GGAs. We intend to test this novel methodology as soon as it becomes available for general usage. The failure of meta-GGAs in reproducing the correct magnetic ground state stems from their tendency to overestimate magnetism, which manifests in larger exchange splittings and local magnetic moments. For SCAN, this effect can be partially mitigated by the process of deorbitalization that leads to SCAN-L, but the latter still performs worse than GGA. Overall, an overestimated magnetism favors extended c-AF regions and leads to an increased magnetic frustration around the SDW nodes, resulting in more rectangular SDW profiles. The associated increase in frustration energy at the nodal sites further decreases the energetic stability of the SDW state. These observations are consistent with earlier reports of overmagnetization in $3d$ transition metals by meta-GGA functionals and highlight the subtle energetic competition that governs magnetic behavior of Cr. As a final comment, the relaxed interplanar distances for SDW cells did not play any significant role in improving the energetics, thereby validating the conclusions of previous studies \cite{hafner}.\\
\noindent Before concluding, we should also acknowledge two methodological limitations. First, we note the use of standard PAW datasets from VASP, which were originally created for the PBE-GGA functional: even when using more advanced meta-GGA functionals, the core electrons are treated as if PBE were applied \cite{Mejia_Rodriguez_2018}. This is a common and accepted practice, and previous studies (using SCAN in particular \cite{Fu_2018_applicability_of_SCAN,Fu_2019,Ekholm_2018_assessing_the_SCAN_functional}) indicate that the impact is generally minor \cite{analysis_of_overmagn}. However, the magnetostructural properties of Cr have been shown to be sensitive to variations in core-electron treatment \cite{barreteau}, so potential side effects cannot be completely discounted. Second, as already mentioned, we have used the pre-converged wavefunctions as starting point to facilitate the meta-GGA calculations. Although our test calculations showed that this strategy should not cause any bias for the lattice parameters used to model the SDW formation, unexpected consequences cannot be ruled out. Nevertheless, the scale of the calculated energy differences makes this scenario unlikely. In summary, although technically more advanced, meta-GGA functionals yield inferior results for bcc Cr if compared to the PBE-GGA. Future work should explore whether additional physical effects — such as lattice distortions, non-local exchange–correlation potentials, or beyond-DFT methods — are needed to accurately describe the SDW ground state in Cr, whose long-range, collective aspects may evade a quasi-local perspective.
 
 
\section*{Acknowledgments}\vspace{-0.5cm}
 \noindent Computational work was mainly performed on the Prime Cluster at the Institute of Physics, Nicolaus Copernicus University, and on resources provided by the National Academic
Infrastructure for Supercomputing in Sweden (NAISS) which are funded in part by the Swedish Research Council through grant agreement
no.~2022-06725. We also gratefully acknowledge the Polish high-performance computing infrastructure PLGrid (HPC Centers: WCSS, ACK Cyfronet AGH) for providing computer facilities and support within the computational grant no.~PLG/2023/016367. This research was carried out as part of the TSSP ExSci–Toruń Students Summer Program in Exact Sciences 2024, and also received some financial support from this program. The authors thank Szymon Śmiga and Oscar Grånäs for technical help and support. The authors would also like to thank the organizers of the TSSP ExSci program for their support and availability. I.D.M. acknowledges financial support from the European Research Council (ERC), Synergy Grant FASTCORR, Project No.~854843, as well as from the STINT Mobility Grant  for Internationalization (Grant No.~MG2022-9386).
\section*{Declaration of generative AI and AI-assisted technologies in the writing process} \vspace{-0.5cm}
\noindent During the preparation of this work the authors used ChatGPT (OpenAI) in order to improve clarity of the manuscript text. After using this tool, the authors reviewed and edited the content as needed and take full responsibility for the content of the published article.


\end{document}



\title{Supplementary: A first-principles study of bcc chromium beyond the generalized gradient approximation (GGA)}
\author{Alma Partos}
\email{partos@doktorant.umk.pl}
\affiliation{Institute of Physics, Nicolaus Copernicus University, 87-100 Toru\'n, Poland}
\author{Igor Di Marco}
\affiliation{Institute of Physics, Nicolaus Copernicus University, 87-100 Toru\'n, Poland}\affiliation{Department of Physics and Astronomy, Uppsala University, Uppsala 751 20, Sweden}
\author{Shivalika Sharma}
\email{shivalikasharma26@gmail.com}
\affiliation{Institute of Physics, Nicolaus Copernicus University, 87-100 Toru\'n, Poland}
\affiliation{Center for Theoretical Physics of Complex Systems, Institute for Basic Science (IBS), Daejeon 34126, South Korea}
\date{December 2025}
\maketitle
\section{Lattice-parameter relaxation using initial wave functions pre-converged with GGA}\label{appendix: faulty calculations with pre-converged wf}

\begin{figure*}[h!]
    \centering
    \includegraphics[width=\linewidth]{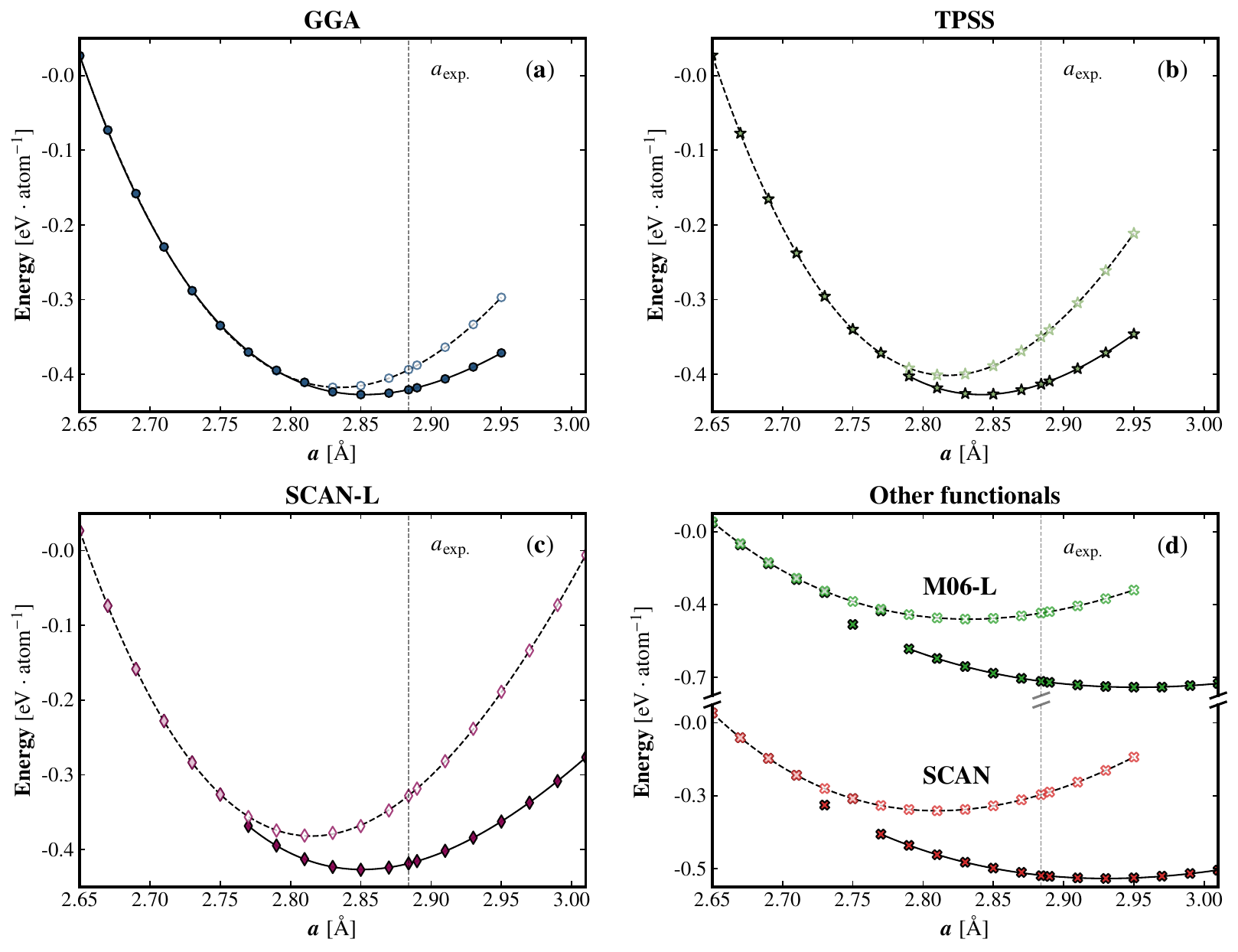}
    \caption{Energy vs.~lattice constant for (a) GGA (b) TPSS, (c) SCAN-L, (d) M06-L and SCAN, using initial wave functions pre-converged in the GGA. Discontinuities are visible for lattice constants around $a \approx 2.79$~Å. This threshold corresponds to the branching point between the NM and AF solutions observed in the GGA [panel (a)].}
    \label{fig:enter-label}
\end{figure*}
\newpage

\section{Local magnetic moments obtained with wave functions pre-converged in the GGA}\label{appendix: faulty calculations with pre-converged wf}
\begin{figure}[h!]
    \centering
    \includegraphics[width=0.5\linewidth]{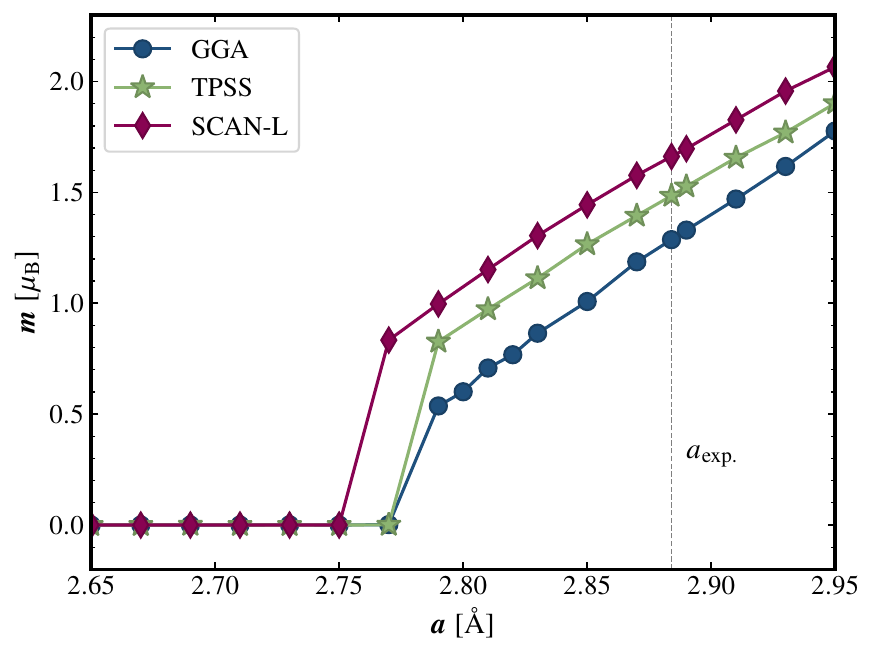}
    \caption{Local magnetic moments as a function of the lattice constant, using initial wave functions pre-converged with the GGA. The magnetization in the meta-GGA formulation is spuriously suppressed if those wave functions are obtained for a GGA ground state that is non-magnetic.}
    \label{fig:enter-label}
\end{figure}

\newpage

\section{Lattice distortions}

\begin{table}[h!]
    \centering
    \caption{Change in average interlayer distance at node and belly sites of the SDW supercell, as obtained from ionic relaxation using GGA and meta-GGA functionals. Deviations are presented as the difference in percent between the relaxed interlayer distances and the corresponding theoretical lattice constant $a_{\text{theo.}}$, for the respective functionals [Table I of the main manuscript]. Here, the node atom is the first atom with a zero-valued magnetic moment, and the belly atom is the central atom, as seen in Figure 1 of the main manuscript. The post-relaxation distance is calculated as the difference between the $\hat{z}$ coordinates of the nearest neighbors of the specified atom $i$, so that $a_i^{\text{relax}} = R_{i + 1} - R_{i - 1}$. We then calculate the fractional deviation as $(a_i^{\text{relax}} - a_{\text{theo.}})/a_{\text{theo.}}$}.
   \setlength{\tabcolsep}{6pt}
   \begin{NiceTabular}{c cS[round-mode=places, round-precision=2]S[round-mode=places, round-precision=2]S[round-mode=places, round-precision=2]S[round-mode=places, round-precision=2]S[round-mode=places, round-precision=2]S[round-mode=places, round-precision=2]}
    \CodeBefore
    \rowlistcolors{3}{white, gray!10}[restart,cols={2-8}]
    \Body 
    \toprule[1pt]
    \Block[draw=black, line-width=1pt]{8-7}{}
    &   & & \multicolumn{5}{c}{\textbf{Deviation from} $\boldsymbol{a_{\text{theo.}}}$ [\%]}   \\
    \cmidrule[1pt](lr){3-7}
      \textbf{Site} & $\boldsymbol{E_{\text{xc}}}$ & {$q/a^*=11/12$} & {$q/a^*=13/14$} & {$q/a^*=17/18$} & {$q/a^*=19/20$} &  {$q/a^*=21/22$}  \\
    \midrule[1pt]
    \Block[draw=black, line-width=0.5pt]{6-1}{}
    \Block[draw=black, line-width=0.5pt]{6-2}{}
    \Block[draw=black, line-width=0.5pt]{6-7}{}
         \multirow{3}{*}{\textbf{Node}} & GGA & -0.34081293894277565 & -0.41671788260076426 & -0.46816719696649045 & -0.4994997923465547 & -0.4953362667311012 \\
           & TPSS & -0.6139761456243065 & -0.7130449984829283 & -0.7283613928951382 & -0.8097403062551507 & -0.7897409867528182   \\
        & SCAN-L & -0.841659650575366 & -0.8365495290425997 &  -1.2437644029288606 & -0.8806657256947545 &    \\
        \cmidrule[0.6pt](lr){1-2}
        \multirow{3}{*}{\textbf{Belly}} & GGA & 0.26142905650094406 & 0.2757874467965564 & 0.28515592952662416 & 0.27055165509007395 & 0.23679599618142788 \\
        & TPSS & 0.36400114818944473 & 0.33069205117897127 & 0.28214526717090394 & 0.30485094187855566 & 0.2388425548349744     \\
        & SCAN-L& 0.31387680620018443 & 0.24205731174187406 & 0.471781411062409 & 0.0767798227910248 &    \\

        \bottomrule[1pt]
    \end{NiceTabular}

\end{table}
\noindent Table SI shows the SDW-induced lattice distortions relative to the unrelaxed structure for each functional at its equilibrium lattice constant, expressed as percentages of the initial (uniform) interlayer distances. In contrast to the GGA and TPSS functionals, the greatest expansions for SCAN-L are generally not found at the belly atoms, but at intermediate points between the node and belly atoms. This is the case for $q/a^* = 11/12$, $13/14$, and $19/20$. A similar situation starts to develop for TPSS at $q/a^* = 21/22$, and may in both cases be understood in connection with the increasingly rectangular magnetic moment profiles. \newline
\noindent For $q/a* = 17/18$ within SCAN-L, the ionic relaxation converges to a qualitatively distinct pattern, which differs from those obtained at neighboring wave vectors and lacks the characteristic central dip observed elsewhere. This solution appears isolated and may correspond to a competing structural configuration of similar energy. Importantly, this behavior does not affect the energetic trends or the conclusions of the present work, and we therefore do not pursue this case further.
\newpage
\section{Fixed-spin-moment calculations}
\begin{figure}[h!]
    \centering
    \includegraphics[width=0.5\linewidth]{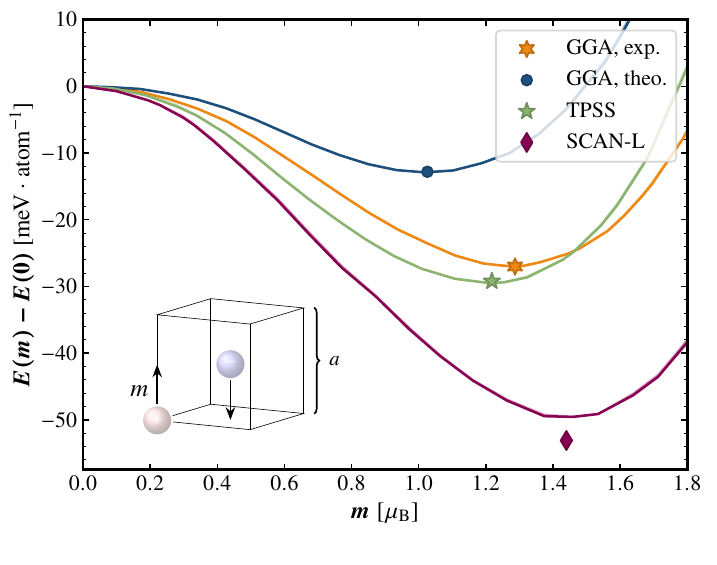}
    \caption{Local magnetic moment versus energy, from fixed-spin moment calculations using the c-AF configuration. The markers indicate the equilibrium local magnetic moments and corresponding total energies, as can be found in Table I of the main material. The TPSS and SCAN-L calculations both use the equilibrium lattice constant.}
    \label{fig:supp_fsm}
\end{figure}

\noindent Fig.~S3 shows the results of constrained fixed-spin-moment (FSM) calculations carried out within the two-atom antiferromagnetic cell. While the residual penalty energies are represented as shaded bands above each curve (enclosing the area between $E(m)$ and $E(m) + E_p(m)$, $E_p(m)$ being the penalty energy for a given moment $m$), these are generally too small to make an immediate visual impression. The energetic cost of enforcing a smaller magnetic-moment magnitude is noticeably larger for the meta-GGA functionals than for the GGA, as reflected by the steeper slopes of the energy curves in the low-moment region (approximately $m \lesssim 0.8$ $\mu_{\text{B}}$). \\
\noindent Note, however, that the SCAN-L curve exhibits a small discrepancy between the minimum total energy estimated via FSM and the total energy of the AF state determined through a standard, unconstrained calculation. This discrepancy amounts to circa $3.6$ meV/atom and likely originates from the technical differences between the two types of calculations. In particular, the FSM calculations require the usage of a non-collinear formalism with constraints on the local magnetic moments, whose definition depends on an arbitrary partition of the physical space. Although previous works have pointed out that FSM calculations with SCAN-based functionals may suffer from numerical errors (\emph{cf.}~Ref.~[39]), the consistency of our FSM results with respect to a different starting point (electron density) suggests that the origin of the observed discrepancy is not numerical but methodological. Despite this issue, the magnetic moment corresponding to the lowest energy is in good agreement with the self-consistent magnetic moment in the AF state. The other functionals do not exhibit this problem and the observed energy differences are within our convergency parameters. Thus, we can reasonably conclude that the relative difficulty of quenching the local magnetic moments follows the order:
$\text{SCAN-L}(a_\text{theo.}) > \text{TPSS}(a_\text{theo.}) > \text{GGA}(a_\text{exp.}) > \text{GGA}(a_\text{theo.})$.